\documentclass[12pt]{article}
\usepackage{amsmath, graphicx}
\usepackage{amstext,amsfonts,amsbsy,eucal,amssymb, amsthm, color}

\theoremstyle{definition}

\newcommand{\be}{\begin{equation}}
\newcommand{\ee}{\end{equation}}
\newcommand{\beqa}{\begin{eqnarray}}
\newcommand{\eeqa}{\end{eqnarray}}
\begin{document}
\begin{center}
{\Large On some new forms of lattice integrable equations}
\end{center}
\begin{center}
Corina N. Babalic$^{1,*}$, A. S. Carstea$^*$
\end{center}
\begin{center}
$^1${\it University of Craiova, Dept. of Physics,\\
13 A.I. Cuza, 200585, Craiova, Romania\\
$^*$National Institute of Physics and Nuclear Engineering, Dept. of Theoretical Physics, Atomistilor 407, 077125, Magurele, Bucharest, Romania\\ }  
\end{center}

\begin{abstract}
Inspired by the forms of delay-Painleve equations, we consider some new differential-discrete systems of KdV, mKdV and Sine-Gordon - type related by simple one way Miura transformations to classical ones. Using Hirota bilinear formalism we construct their new integrable discretizations, some of them having higher order. In particular, by this procedure, we show that the integrable discretization of intermediate sine-Gordon equation is exactly lattice mKdV and also we find a bilinear form of the recently proposed lattice Tzitzeica equation. Also the travelling wave reduction of these new lattice equations is studied and it is shown that all of them, including the higher order ones, can be integrated to Quispel-Roberts-Thomson (QRT) mappings.
\end{abstract}

pacs: {02.30.Ik, 02.30.Ks, 05.45.-a}

\section{Introduction}

Integrable dynamics is a fascinating topic which has progressed and developed rapidly in the last decades. Roughly speaking, it is associated with the existence of a huge amount of internal or hidden symmetry and this fact explains quite well the predictability and regularity which characterize the integrable systems. Lattice equations (partial {\it difference} or P$\Delta$E's),  which exhibit a higher complexity, attracted many studies in the last years \cite{all0-4}--\cite{all0-3}. The main progress was possible due to appearance of some efficient tools in proving integrability such as singularity confinement \cite{all1, RG}, "cube consistency" \cite{all2-0}--\cite{all2-2} and complexity growth \cite{all3-0}--\cite{seb}.
  
Intensively studied in the past, the lattice versions of classical soliton equations have proven to be instrumental in establishing a classification of quadrilateral lattice equations based on some symmetry assumptions (like cube consistency and tetrahedron property)\cite{all2-0}. However, in the case of higher order equations the situation is different. Because of the higher complexity they are studied mainly by examples, although in principle, from the Sato theory and Miwa transformations \cite{sato}--\cite{sato4}, one can obtain lattice equations corresponding to a given hierachy. 
However there is another way to construct lattice equations starting directly from some differential-difference integrable versions and discretize directly by the Hirota bilinear method. It is easy to discretize the bilinear form but  problems may appear at re-obtaining the nonlinear form. The integrability is proved by showing the existence of {\it three soliton solution}. The construction of 1-soliton and 2-soliton solution is easy in  bilinear form and can also be done for some non-integrable systems. But  the 3-soliton solution imposes strong restrictions on the equations and, accordingly is practically equivalent with integrability. It was used succesfully for classifying integrable partial differential equations \cite{hietarinta-all1}--\cite{hietarinta-all4} and quite recently for  P$\Delta$E's on a $3\times 3$ stencil \cite{jarmo-hm}.
In this paper we are constructing some new lattice versions of KdV and mKdV equations starting from new differential-difference forms related to delay-Painleve equations. We use Hirota bilinear formalism for both construction and proving integrability. Also, starting from the bilinear form of semidiscrete sine-Gordon equation we find again the recently proposed lattice Tzitzeica equation. In the last section we make the travelling wave reduction of the above lattice equations and we show that all of them can be integrated to classical QRT mappings \cite{qrt}

\section{New integrable lattice KdV equations}

We start from a new differential-difference equation namely:
\begin{equation}
\label{eq1}
\dot w_n=\frac{w_n}{w_{n-1}}(w_{n+1}-w_{n-1}).
\end{equation}

The travelling wave reduction of this equation is exactly the autonomous limit of three point delay-Painleve I equation \cite{alfreddel}, \cite{fanedel}. There is also a one way Miura transformation from (\ref{eq1}) to the well known differential-difference KdV equation of Hirota \cite{hirsat};  if $w_n$ is a solution of (\ref{eq1}), then $u_n=w_n/w_{n-1}$ is a solution of 
\begin{equation}
\label{eqq}
\dot u_n-u_n^2(u_{n+1}-u_{n-1})=0.
\end{equation}

Indeed if we put $u_n=w_n/w_{n-1}$, we obtain (after rearranging terms):
$$w_{n-2}({\dot w_n}w_{n-1}-w_nw_{n+1})-w_n({\dot w_{n-1}}w_{n-2}-w_nw_{n-1})=0.$$

This relation is true since both brackets can be computed from  (\ref{eq1}) and its downshift along $n$. 

Integrability of (\ref{eq1}) can be proved by Hirota bilinear formalism. If  $w_n=F_{n+1}(t)/F_n(t)$, then the following bilinear system is found (for notational simplicity we remove $t$-dependence):

\begin{equation}
\label{eq4}
D_tF_{n+1}\cdot F_n=F_{n+2}F_{n-1}-F_nF_{n+1},
\end{equation}
which is exactly the bilinear form of the classical integrable differential-difference KdV equation (\ref{eqq}) \cite{hirotakdv}. This fact shows that the equation (\ref{eq1}) is a potential semidiscrete KdV equation. 

In order to discretize a bilinear system  we have to discretize the Hirota operator  in (\ref{eq4}) by replacing time derivatives with finite differences with the step $\delta$ (and $t\to \delta m$):
$$D_tG_n\cdot F_n\to\left(\frac{1}{\delta}[G(n,(m+1)\delta)-G(n,\delta m)]\right)F(n,\delta m)-$$
$$-\left(\frac{1}{\delta}[F(n,(m+1)\delta)-F(n,\delta m)]\right)G(n,\delta m)$$

After that, we impose the invariance of the resulting bilinear equations with respect to multiplication with $\exp(\mu n+\nu m)$ for any $\mu, \nu$ (the bilinear gauge invariance).

The discretization of (\ref{eq4}) is straitforward. We obtain the classical reduced Hirota-Miwa equation:
$$F_{n+1}^{m+1}F_n^m-F_{n+1}^mF_n^{m+1}=\delta[ F_{n+2}^{m+1}F_{n-1}^m- F_{n+1}^mF_n^{m+1}].$$

Consider $W_n^m=F_{n+1}^m/F_n^m$. Then, dividing the bilinear equation with $F_n^{m+1}F_n^m$, we obtain the following:
\begin{equation}
\label{eq14}
W_n^{m+1}=(1-\delta)W_n^m+\delta\frac{W_{n+1}^{m+1}W_{n}^{m+1}}{W_{n-1}^m},
\end{equation} which is a quad-lattice equation and it is the discrete analogue of (\ref{eq1}).  One can easily see that it is not consistent around the cube so it is not in the ABS-list. To our knowledge, this is a new equation and, parallel to the semidiscrete case, there is a simple Miura transformation from it to classical lattice KdV equation of Hirota \cite{hirotakdv}. More precisely, if $W_n^m$ is a solution of (\ref{eq14}), then:
\begin{equation}
\label{eq15}
u_n^m=W_n^{m+1}/W_{n-1}^m
\end{equation} 
obeys the lattice KdV of Hirota:
\begin{equation}
\label{eq16}
u_n^{m+1}-u_n^m=\frac{\delta}{1-\delta}u_n^{m+1}u_n^m(u_{n+1}^{m+1}-u_{n-1}^m).
\end{equation}

Indeed, if $u_n^m=W_n^{m+1}/W_{n-1}^m$ then (\ref{eq16}) becomes:
$$W_n^{m+2}\left[(1-\delta)W_{n-2}^mW_{n-1}^m+\delta W_{n-1}^{m+1}W_n^{m+1}\right]-$$
$$-W_{n-2}^m\left[(1-\delta)W_{n}^{m+1}W_{n-1}^{m+1}+\delta W_{n}^{m+2}W_{n+1}^{m+2}\right]=0,$$
which is true since the first square bracket can be computed from one downshift along the $n$-direction of (\ref{eq14}) and the second square bracket from the one upshift along the $m$-direction.

\section{New integrable lattice form of mKdV equation}

In this section we are starting from the following new equation:
\begin{equation}
\label{eq17}
\dot v_n=2v_n\frac{v_{n+1}-v_{n-1}}{v_{n+1}+v_{n-1}},
\end{equation}
which goes again in the travelling wave reduction to the three-point autonomous delay-Painleve II \cite{alfreddel}. Also, there is a Miura from (\ref{eq17}) to  the classical semidiscrete mKdV equation (self-dual nonlinear network) \cite{hs1, hirsat}:
\begin{equation}\label{eq17a}
\dot u_n=(1+u_n^2)(u_{n+1}-u_{n-1}),
\end{equation}
namely, $u_n=\frac{i}{2}\frac{d}{dt}\log v_n$. This fact can be seen immediately since (\ref{eq17}) is equivalent with:
$$\frac{i}{2}\frac{d}{dt}\log v_n=\tan\left(\frac{i}{2}\log{v_{n+1}}-\frac{i}{2}\log{v_{n-1}}\right).$$

Applying $\tan^{-1}$ in both members and then taking the derivative with respect to $t$, one gets exactly (\ref{eq17a}). So the equation (\ref{eq17}) is a potential new semidiscrete mKdV equation.

We build the Hirota bilinear form using the substitution $v_n=G_n(t)/F_n(t)$. Introducing in (\ref{eq17}) and decoupling in the bilinear dispersion relation and the soliton-phase constraint, we obtain the following system:
\begin{eqnarray}
\label{eq18}
D_t G_n\cdot F_n&=&G_{n+1}F_{n-1}-G_{n-1}F_{n+1},\\
\label{eq19}
2G_nF_n&=&G_{n+1}F_{n-1}+G_{n-1}F_{n+1}.
\end{eqnarray}

The above system is an integrable one since it admits 3-soliton solution of the following form ($k_i$ is the wave number, $\omega_i$ is the angular frequency):
\begin{eqnarray}
\label{eq20}
F_n=\sum_{\mu_1,\mu_2,\mu_3\in \{0,1\}}\left(\prod_{i=1}^3 (a_i e^{\eta_i})^{\mu_i}\right)\prod_{i<j}^3 A_{ij}^{\mu_i\mu_j},\\
\label{eq21}
G_n=\sum_{\mu_1,\mu_2,\mu_3\in \{0,1\}}\left(\prod_{i=1}^3 (b_i e^{\eta_i})^{\mu_i}\right)\prod_{i<j}^3 A_{ij}^{\mu_i\mu_j},
\end{eqnarray}
where $\eta_i=k_i n+\omega_i t,\quad i=\overline {1,3}$.\\

The dispersion relation has the form:
\begin {equation}
\label{eq22}
\omega_i=e^{k_i}-e^{-k_i}=2\sinh k_i.
\end {equation}

Phase factors and the interaction terms are:
\begin{eqnarray}
\label{eq23}
 b_i=-a_i=1,\\
\label{eq24}
A_{ij}=\frac {\cosh(k_i-k_j)-1}{\cosh(k_i+k_j)-1}&,& \quad i<j=\overline {1,3}.
\end{eqnarray} 

We discretize the bilinear system (\ref{eq18})-(\ref{eq19}) in the same way. Replacing time derivatives in (\ref{eq18}) with finite differences and imposing the bilinear gauge invariance, we obtain:
\begin{eqnarray}
\label{eq25}
G_n^{m+1}F_n^m-G_n^mF_n^{m+1}&=&\delta(G_{n+1}^{m+1}F_{n-1}^m-G_{n-1}^mF_{n+1}^{m+1}),\\
\label{eq26}
2G_n^mF_n^m&=&G_{n+1}^mF_{n-1}^m+G_{n-1}^mF_{n+1}^m.
\end{eqnarray}

The above system admits the following 3-soliton solution: 
\begin{eqnarray}
\label{eq27}
F_n^m=\sum_{\mu_1,\mu_2,\mu_3\in \{0,1\}}\left(\prod_{i=1}^3 a_i^{\mu_i}(p_i^n q_i^{m\delta})^{\mu_i}\right)\prod_{i<j}^3 A_{ij}^{\mu_i\mu_j},\\
\label{eq28}
G_n^m=\sum_{\mu_1,\mu_2,\mu_3\in \{0,1\}}\left(\prod_{i=1}^3 b_i^{\mu_i}(p_i^n q_i^{m\delta})^{\mu_i}\right)\prod_{i<j}^3 A_{ij}^{\mu_i\mu_j}
\end{eqnarray}
with the same phase factors and interaction terms as in the differential-difference case, but different dispersion relation:
\begin{equation}
\label{eq29}
q_i=\left(\frac{1-\delta p_i^{-1}}{1-\delta p_i}\right)^{1/\delta}, \quad i=\overline {1,3},
\end{equation}
where  $p_i=e^{k_i}$, $q_i=e^{\omega_i}$, $i=1,2,3$ ($k_i$ is the wave number and $\omega_i$ is the angular frequency).

Now we can recover the nonlinear form. Dividing (\ref{eq25}) by $F_n^mF_n^{m+1}$ and using the following notations $\omega_n^m=\frac{G_n^m}{F_n^m}$, $\Gamma_n^m=\frac{F_{n-1}^mF_{n+1}^{m+1}}{F_n^mF_n^{m+1}}$, we obtain:
$$\omega_n^{m+1}-\omega_n^m=\delta\Gamma_n^m(\omega_{n+1}^{m+1}-\omega_{n-1}^m).$$

From equation (\ref{eq26}) we find: 
$$\frac{(F_n^m)^2}{F_{n+1}^mF_{n-1}^m}=\frac{\omega_{n-1}^m+\omega_{n+1}^m}{2\omega_n^m}.$$

But one can see immediately:
$$\frac{\Gamma_{n+1}^m}{\Gamma_n^m}=\frac{F_{n+2}^{m+1}F_n^{m+1}}{(F_{n+1}^{m+1})^2}\frac{(F_n^m)^2}{F_{n+1}^mF_{n-1}^m}=\frac{2\omega_{n+1}^{m+1}}{\omega_{n+2}^{m+1}+\omega_n^{m+1}}\frac{\omega_{n+1}^m+\omega_{n-1}^m}{2\omega_n^m}.$$

Finally the nonlinear form of our sistem is:
\begin{eqnarray}
\label{eq30}
\omega_n^{m+1}-\omega_n^m&=&\delta\Gamma_n^m(\omega_{n+1}^{m+1}-b\omega_{n-1}^m),\\
\Gamma_{n+1}^m&=&\frac{\omega_n^m+\omega_{n-1}^m}{\omega_{n+2}^m+\omega_n^{m+1}}\frac{\omega_{n+1}^{m+1}}{\omega_n^m}\Gamma_n^m.\nonumber
\end{eqnarray}

One can eliminate $\Gamma_n^m$ and $\Gamma_{n+1}^m$ and we get the following new higher order nonlinear lattice mKdV equation:
\begin{equation}
\label{eq31}
\frac{\omega_n^{m+1}-\omega_n^m}{\omega_{n+1}^{m+1}-\omega_{n+1}^m}=\frac{\omega_{n+1}^{m+1}-\omega_{n-1}^m}{\omega_{n+2}^{m+1}-\omega_n^m}\frac{\omega_{n+2}^{m+1}+\omega_n^{m+1}}{\omega_n^m+\omega_{n+1}^m}\frac{\omega_n^m}{\omega_{n+1}^{m+1}}.
\end{equation}

\section{Lattice system related to intermediate sine-Gordon equation}

In this section we are going to study the following differential-difference equation:
\begin{equation}
\label{eq32}
\frac{d}{dt}(u_{n}u_{n+1})=\gamma u_n^2+\kappa{u_{n+1}}^2,
\end{equation}
where $\gamma$ and $\kappa$ are constants. In the travelling wave reduction it goes to the bi-Riccati form of the delay-Painleve II equation \cite{fanedel}. On the other hand, for $\gamma=-\kappa=1$ the above equation is equivalent with the famous intermediate sine-Gordon equation \cite{matsuno}:
$$\partial_tTy(x,t)+2\sin y(x,t)=0,$$ 
where $T$ is a singular integral operator defined through: 
$$(Tf)(x)=\frac{1}{2\sigma}P\int_{-\infty}^{\infty}\coth{\frac{\pi(z-x)}{2\sigma}}f(z)dz.$$

One of the key properties of the $T$ operator is that it acts as a "jump" among different strips on the complex $x$-plane. So starting from a meromorphic function $u(x,t)$ in a horizontal strip of width $2\sigma$ of the complex $x$-plane, we define $y(x,t)=i\log{\bar{u}/u}$ where $\bar{u}=u(x+i\sigma)$ and $u=u(x-i\sigma)$. Then $Ty=-\log{\bar{u}u}$, so the intermediate sine-Gordon becomes (\ref{eq32}) for $n=(x-i\sigma)/2i\sigma$.

The bilinear form of (\ref{eq32}) can be computed by taking $u_n=G_n/F_n$. Taking into account the following relation (which can be checked by direct computation),
$$G_{n+1}F_{n+1}D_tG_n\cdot F_n+G_nF_nD_tG_{n+1}\cdot F_{n+1}=G_{n+1}F_nD_tG_n\cdot F_{n+1}+G_nF_{n+1}D_tG_{n+1}\cdot F_n$$
the equation (\ref{eq32}) will turn into:
$$G_{n+1}F_n(D_tG_n\cdot F_{n+1}-\kappa G_{n+1}F_n)+G_nF_{n+1}(D_tG_{n+1}\cdot F_n-\gamma G_n F_{n+1})=0.$$

Now it can be splitted in the following way:
\begin{equation}
\label{eq33}
D_t G_n\cdot{F_{n+1}}-\kappa{G_{n+1}}F_n=AG_nF_{n+1},
\end{equation}
\begin{equation}
\label{eq34}
D_t {G_{n+1}}\cdot{F_n}-\gamma G_n{F_{n+1}}=-A{G_{n+1}}{F_n},
\end{equation}
where A is a gauge constant. Replacing time derivatives with finite differences and  $t$ with $\delta m$ in (\ref{eq33}) and (\ref{eq34}),  and imposing the bilinear gauge invariance, we contruct the fully discrete gauge invariant bilinear equations:
\begin{eqnarray}
\label{eq35}
G_n^{m+1}F_{n+1}^m-G_n^mF_{n+1}^{m+1}&=&\delta(kG_{n+1}^{m+1}F_n^m+AG_n^mF_{n+1}^{m+1}),\\
\label{eq36}
G_{n+1}^{m+1}F_n^m-G_{n+1}^mF_n^{m+1}&=&\delta(\gamma G_n^mF_{n+1}^{m+1}-AG_{n+1}^{m+1}F_n^m).
\end{eqnarray}

We take A=1, $\gamma=-\kappa=1$ and the above system admits the following $3$-soliton solution: 
\begin{eqnarray}
\label{eq37}
F_n^m=\sum_{\mu_1,\mu_2,\mu_3\in \{0,1\}}\left(\prod_{i=1}^3 a_i^{\mu_i}(p_i^n q_i^{m\delta})^{\mu_i}\right)\prod_{i<j}^3 A_{ij}^{\mu_i\mu_j},\\
\label{eq38}
G_n^m=\sum_{\mu_1,\mu_2,\mu_3\in \{0,1\}}\left(\prod_{i=1}^3 b_i^{\mu_i}(p_i^n q_i^{m\delta})^{\mu_i}\right)\prod_{i<j}^3 A_{ij}^{\mu_i\mu_j},
\end{eqnarray}
where the dispersion relation is:
$$q_i=\frac{1+2\delta+p_i}{1+(1+2\delta)p_i}.$$

The phase factors are:
$$a_i=-b_i=1, \quad i=\overline{1,3}$$
and the interaction terms: 
$$A_{ij}=\left(\frac{p_i-p_j}{1-p_ip_j}\right)^2.$$

In order to see the nonlinear form, we take $X_n^m=G_n^m/F_n^m$ and from bilinear equation we obtain (after eliminating the term $F_n^{m+1}F_{n+1}^m/F_n^mF_{n+1}^{m+1}$):
$$X_{n+1}^m=X_{n}^{m+1}\frac{X_{n+1}^{m+1}(1+\delta)-\delta X_{n}^m}{X_{n}^{m}(1+\delta)-\delta X_{n+1}^{m+1}},$$
which is nothing but classical lattice mKdV equation \cite{frank}. So, we can say that the lattice version of the polynomial intermediate sine-Gordon equation is the classical lattice mKdV. It is well known that lattice mKdV and lattice sine-Gordon are practically equivalent \cite{alfred-sg} (up to a exponentiation of the dependent variable with $(-1)^m$), so now we can say that lattice intermediate sine-Gordon is equivalent with lattice sine-Gordon (up to the above mentioned exponentiation). \\
\noindent {\bf Remark 1:}\\
{ \it Remark on discrete Tzitzeica equation:} An interesting, although formal connection can be made with the discrete Tzitzeica equation which has been introduced quite recently by Adler \cite{adler}. Starting from the general differential-difference equation (which goes to various forms of bi-Riccati delay-Painleve III in the travelling wave reduction):
\begin{equation}
\label{eq39}
\frac{\bar{u}'}{\bar{u}}-\frac{u'}{u}=\alpha u\bar{u}+\frac{\mu}{u\bar{u}}+\beta_0 u+{\beta_1}\bar{u}
\end{equation}
we get by means of $u=G/F$:
\begin{equation}
\label{eq40}
D_t \bar{G}\cdot{G}-\mu F\bar{F}=A\bar{G}G,
\end{equation}
\begin{equation}
\label{eq41}
D_t \bar{F}\cdot{F}+\alpha\bar{G}G+\beta_0 G\bar{F}+\beta_1\bar{G}F=AF\bar{F},
\end{equation}
where A is a constant. 

We discretize in a gauge invariant way the above bilinear form and we get the following general bilinear system:
\begin{equation}
\label{eq42}
G_{n+1}^{m+1}G_n^m-G_{n+1}^mG_n^{m+1}-\delta \mu F_n^{m+1}F_{n+1}^m=\delta A G_{n+1}^mG_n^{m+1},
\end{equation}
\begin{equation}
\label{eq43}
F_{n+1}^{m+1}F_n^m-F_{n+1}^mF_n^{m+1}+\delta\alpha G_{n+1}^{m}G_n^{m+1}+\delta\beta_{00} G_n^{m+1}F_{n+1}^m+$$
$$+{\delta\beta_{10}} G_{n+1}^mF_n^{m+1}+\delta\beta_{01} G_n^mF_{n+1}^{m+1}+{\delta\beta_{11}} G_{n+1}^{m+1}F_n^m=\delta AF_{n+1}^mF_n^{m+1},
\end{equation}
where $\beta_{00},\beta_{01}, \beta_{10},\beta_{11}$ are arbitrary coefficients, which have to be determined according to integrability requirements.

Now consider $\delta A=-1, \delta\mu=1, \delta\alpha=-1/c^2, \beta_{00}=-\beta_{01}=1/{\delta c}, \beta_{10}=-\beta_{11}=1/{\delta c}$ (this choice of coefficients will make more difficult the continuum limit $\delta\to 0$). Then the above bilinear system will have the form:
\begin{equation}
\label{eq44}
G_{n+1}^{m+1}G_n^m-F_n^{m+1}F_{n+1}^m=0,
\end{equation}
\begin{equation}
\label{eq45}
cF_{n+1}^{m+1}F_n^m-\frac{1}{c}G_{n+1}^{m}G_n^{m+1}+G_n^{m+1}F_{n+1}^m+G_{n+1}^mF_n^{m+1}-G_n^mF_{n+1}^{m+1}-G_{n+1}^{m+1}F_n^m=0.
\end{equation} 

Calling $W_n^m=G_n^m/F_n^m
$ and eliminating the term $F_n^mF_{n+1}^{m+1}/F_{n+1}^mF_n^{m+1}$, we will get exactly the form of lattice Tzitzeica equation found by Adler \cite{adler}:
$$\frac{W_n^mW_{n+1}^{m+1}}{c-W_n^m-W_{n+1}^{m+1}}=\frac{1}{c^{-1}W_{n+1}^mW_n^{m+1}-W_n^{m+1}-W_{n+1}^m}$$

We claim that the bilinear system (\ref{eq44}), (\ref{eq45}) is the Hirota bilinear form of the lattice Tzitzeica equation. Indeed, if we put $G_n^m=\tau_{n+1}^m\tau_{n}^{m+1}, F_n^m=\tau_{n}^m\tau_{n+1}^{m+1}$, then the first bilinear equaton is identically verified and the second bilinear equation is (up to a common factor) exactly the trilinear form found by Adler \cite{adler}:
$$
\det\left(
\begin{array}{ccc}
\tau_{n}^{m+2}&\tau_{n+1}^{m+2}&\tau_{n+2}^{m+2}\\
\tau_{n}^{m+1}&c^{-1}\tau_{n+1}^{m+1}&\tau_{n+2}^{m+1}\\
\tau_n^m&\tau_{n+1}^m&\tau_{n+2}^m\\
\end{array}\right)-(c-c^{-1})\tau_n^m\tau_{n+1}^{m+1}\tau_{n+2}^{m+2}=0.
$$

\section{Reductions of lattice equations}

\subsection{ Reduction of the potential lattice KdV}

In this section we are going to study various $(p,q)$-reductions or travelling wave reductions of the above lattice equations. We will show that even though some lattice equations are of higher order, their travelling wave reductions can be integrated to classical QRT-mappings \cite{qrt}. The interesting fact is that both additive and multiplicative QRT-mappings are obtained (additive and multiplicative refers to the type of rational surface on which the mappings act as automorphisms \cite{sakai}). The general form of a symmetric QRT mapping is the following:
$$x_{m+1}=\frac{f_1(x_m)-x_{m-1}f_2(x_m)}{f_2(x_m)-x_{m-1}f_3(x_m)}$$
where $f_1,f_2,f_3$ are general quartic polynomials in $x_m$. Any QRT mapping possesses an invariant which is biquadratic in $x_m$ and $x_{m-1}$. The integrability comes from the fact that this biquadratic correspondence can be integrated in terms of elliptic functions (result which goes as far back as Euler).

Let us start with the quad-lattice KdV equation (\ref{eq14}):
$$
W_n^{m+1}=(1-\delta)W_n^m+\delta\frac{W_{n+1}^{m+1}W_{n}^{m+1}}{W_{n-1}^m}.$$

We consider that $W_n^m=x(n+m)\equiv x_{\nu}$ with $\nu=n+m$. In this reduction our equation becomes the following  mapping:
$$x_{\nu+1}x_{\nu-1}-(1-\delta)x_\nu x_{\nu-1}-\delta x_{\nu+2}x_{\nu+1}=0.$$

Dividing by $x_{\nu+1}x_{\nu}$ we get:
$$\frac{x_{\nu-1}}{x_{\nu}}-(1-\delta)\frac{x_{\nu-1}}{x_{\nu+1}}-\delta \frac{x_{\nu+2}}{x_{\nu}}\underbrace{=}_{w_\nu=x_{\nu+1}/x_{\nu}}$$
$$=1/w_{\nu-1}-(1-\delta)/(w_{\nu}w_{\nu-1})-\delta w_{\nu+1}w_\nu\underbrace{\Leftrightarrow}_{\alpha=-1+1/\delta}$$
$$w_{\nu+1}w_{\nu-1}=\frac{\alpha+1}{w_\nu}-\frac{\alpha}{w_{\nu}^2}.$$

The last relation is a QRT mapping which is the autonomous limit of a $q$-Painleve equation realizing an automorphism of a rational surface of type $A_{7}^{(1)}$\cite{sakai}.

\noindent{\bf Remark2:} The travelling wave reduction of classical lattice KdV of Hirota gives a QRT mapping which is the autonomous limit the $d$-Painleve equation (aditive type of rational surface $E_6^{(1)}$)\cite{sakai}. 
It was shown in \cite{reduc-fane} that, if $u_n^m=u(n+m)\equiv u_{\nu}, \delta'=\delta/(1-\delta)$, then we get from (\ref{eq16}):
$$u_{\nu+1}+u_{\nu}+u_{\nu-1}=\gamma-\frac{1}{\delta'}\frac{1}{u_\nu}$$

What is really interesting is that the two equations (\ref{eq14}) and  (\ref{eq16}) give, by the same reduction, both multiplicative and additive QRT mappings.

\subsection{Reduction of the higher order lattice mKdV}

The higher lattice mKdV is (\ref{eq31}):
$$\frac{\omega_n^{m+1}-\omega_n^m}{\omega_{n+1}^{m+1}-\omega_{n+1}^m}=\frac{\omega_{n+1}^{m+1}-\omega_{n-1}^m}{\omega_{n+2}^{m+1}-\omega_n^m}\frac{\omega_{n+2}^{m+1}+\omega_n^{m+1}}{\omega_{n-1}^m+\omega_{n+1}^m}\frac{\omega_n^m}{\omega_{n+1}^{m+1}}.$$

We consider that $\omega(n,m)=x(n+m)\equiv x_{\nu}$ with $\nu=n+m$. In this reduction our equation becomes the following four-order mapping:
$$\frac{x_{\nu+1}-x_\nu}{x_{\nu+2}-x_{\nu+1}}-\frac{x_{\nu+2}-x_{\nu-1}}{x_{\nu+3}-x_\nu}\frac{x_{\nu+3}+ x_{\nu+1}}{x_{\nu-1}+x_{\nu+1}}\frac{x_\nu}{x_{\nu+2}}=0.$$

Then we have:
$$\frac{x_{\nu+1}-x_\nu}{x_{\nu+2}-x_{\nu+1}}-\frac{x_{\nu+2}-x_{\nu-1}}{x_{\nu+3}-x_\nu}
\frac{x_{\nu+3}+ x_{\nu+1}}{x_{\nu-1}+x_{\nu+1}}\frac{x_\nu}{x_{\nu+2}}=$$
$$=\frac{x_{\nu+1}/x_\nu-1}{x_{\nu+2}/x_{\nu+1}-1}\frac{x_\nu}{x_{\nu+1}}-\frac{x_{\nu+2}/x_{\nu-1}-1}{x_{\nu+3}/x_\nu-1}\frac{x_{\nu+3}/x_{\nu+1}+1}{x_{\nu+1}/x_{\nu-1}+1}\frac{x_{\nu+1}}{x_{\nu+2}}=$$
$$=\frac{x_{\nu+1}/x_\nu-1}{x_{\nu+2}/x_{\nu+1}-1}\frac{x_\nu}{x_{\nu+1}}-
\frac{(x_{\nu+2}/x_{\nu+1})(x_{\nu+1}/x_{\nu})(x_{\nu}/x_{\nu-1})-1}{(x_{\nu+3}/x_{\nu+2})(x_{\nu+2}/x_{\nu+1})(x_{\nu+1}/x_\nu)-1}\times$$
$$\times\frac{(x_{\nu+3}/x_{\nu+2})(x_{\nu+2}/x_{\nu+1})+1}{(x_{\nu+1}/x_{\nu})(x_{\nu}/x_{\nu-1})+1}\frac{x_{\nu+1}}{x_{\nu+2}}\underbrace{=}_{w_\nu=\frac{x_{\nu+1}}{x_{\nu}}}$$
$$\frac{w_\nu-1}{w_{\nu+1}-1}-\frac{w_{\nu+1}w_{\nu}w_{\nu-1}-1}{w_{\nu+2}w_{\nu+1}w_\nu-1}\frac{w_{\nu+2}w_{\nu+1}+1}{w_{\nu}w_{\nu-1}+1}\frac{w_{\nu}}{w_{\nu+1}}=0.$$

Now we force $(w_\nu w_{\nu+1}+1)$ as a denominator in the second term. We get
$$\frac{w_\nu-1}{w_{\nu+1}-1}-\frac{w_{\nu+1}w_{\nu}w_{\nu-1}-1}{w_{\nu+2}w_{\nu+1}w_\nu-1}\frac{w_{\nu+2}w_{\nu+1}+1}{w_{\nu+1}w_{\nu}+1}\frac{w_{\nu+1}w_{\nu}+1}{w_{\nu}w_{\nu-1}+1}\frac{w_{\nu}}{w_{\nu+1}}=0\Longleftrightarrow$$
$$\Longleftrightarrow \frac{(w_{\nu}-1)(w_{\nu+1} w_\nu+1)(w_\nu w_{\nu-1}+1)}{(w_{\nu+1}w_\nu w_{\nu-1}-1)w_{\nu}}=\frac{(w_{\nu+1}-1)(w_{\nu+2}w_{\nu+1}+1)(w_{\nu+1} w_{\nu}+1)}{(w_{\nu+2}w_{\nu+1}w_{\nu}-1)w_{\nu+1}}.$$

Again the left hand side is the downshift of the right hand side, so both members are equal to a constant $\sigma$. 
Accordingly, the integrated reduction of the lattice mKdV is the following more complicated QRT-form:
$$w_{\nu+1}=\frac{1-w_{\nu}(\sigma+1)+w_{\nu-1}(w_{\nu}-w_{\nu}^2)}{-(w_{\nu}-w_{\nu}^2)-w_{\nu-1}(w_{\nu}^2(1+\sigma)-w_{\nu}^3)}.$$

\noindent {\bf Remark3:} Travelling wave reduction of Tzitzeica equation gives trivially the QRT mapping:
$$w_{\nu+1}w_{\nu-1}=\frac{c-w_{\nu+1}-w_{\nu-1}}{c^{-1}w_\nu^2-2w_{\nu}}.$$

\section{Conclusions}
In this paper we have considered some differential-difference nonlinear equations related to corresponding integrable nonlinear dynamical systems well known in mathematical physics like KdV, mKdV and sine-Gordon equations. For each of them, we have derived the Hirota bilinear form and introduced a Miura transformation that changes the new semidiscrete equations in already known differential-difference equations. Using Hirota bilinear formalism we have constructed their new integrable discretizations, some of them having higher order. The Hirota bilinear formalism was also used for proving integrability. The paper also shows that the integrable discretization of intermediate sine-Gordon equation is the lattice mKdV, and proposes a bilinear form of lattice Tzitzeica equation \cite{adler}. The travelling wave reduction of all these lattice equations leads to classical QRT mappings.\\

%\begin{acknowledgements}
\noindent {\bf Acknowldgement:} The authors have been supported by the projects PN-II-ID-PCE-2011-50/2011 and PN-II-ID-PCE-2011-3-0083, Romanian Ministery of Education and Research. Corina N. Babalic also acknowledges the support of the grant  FP7-PEOPLE-2012-IRSES-316338. 
%\end{acknowledgements}

\end{document}